\documentclass[aps,pre,onecolumn,epsf,graphicx]{revtex4}
\def\pds{{\partial \over \partial s}}
\def\pdx{{\partial \over \partial x}}
\def\pdxi{{\partial \over \partial \xi}}
\def\pd2x{{\partial^2 \over \partial x^2}}

\usepackage{epsfig,latexsym}
\def\btab{\begin{table}}
\def\btabu{\begin{tabular}}
\def\etab{\end{table}}
\def\etabu{\end{tabular}}
\newcommand \bew {\begin{widetext}}
\newcommand \enw {\end{widetext}}

\begin{document}


\title{\bf\noindent Equilibrium statistics of a slave estimator 
in Langevin processes }

\author{David S. Dean$^{(1,2)}$~, Ian T. Drummond$^{(1)}$~,  
Ron R. Horgan$^{(1)}$
and Satya N. Majumdar$^{(2,3)}$}

\affiliation{
(1) DAMTP, CMS, University of Cambridge, Cambridge, CB3 0WA, UK \\
(2) Laboratoire de Physique Th\'eorique,  UMR CNRS 5152, IRSAMC, Universit\'e 
Paul Sabatier, 118 route de Narbonne, 31062 Toulouse Cedex 04, France\\
 (3) Laboratoire de Physique Th\'eorique et Mod\`eles Statistiques, UMR 8626,
Universit\'e Paris Sud, B\^at 100, 91045 Orsay Cedex, France}

\date{21st September  2004}
\begin{abstract}
We analyze the statistics of an estimator, denoted by $\xi_t$
and referred to as the slave, for the equilibrium susceptibility of a 
one dimensional Langevin process $x_t$ in a  potential $\phi(x)$~. 
The susceptibility can be measured by evolving the slave equation in
conjunction with the original Langevin process. This procedure yields
a direct estimate of the susceptibility and avoids the need, when
performing numerical simulations, to include applied external fields
explicitly.  The success of the method however depends on the
statistical properties of the slave estimator. The joint probability
density function for $x_t$ and $\xi_t$ is
analyzed. In the case where the potential of the system has a concave
component the probability density function of the slave acquires a
power law tail characterized by a temperature dependent exponent. Thus
we show that while the average value of the slave, in the equilibrium
state, is always finite and given by the fluctuation dissipation
relation, higher moments and indeed the variance may show divergences.
The behavior of the power law exponent is analyzed in a general
context and it is calculated explicitly in some specific examples.  Our
results are confirmed by numerical simulations and we discuss possible
measurement discrepancies in the fluctuation dissipation relation
which could arise due to this behavior.
\end{abstract}
\maketitle
\vspace{.2cm} \pagenumbering{arabic}

\section{Introduction}
A standard experimental technique for probing a system is to measure
its response to a small external field. In equilibrium, static response
functions are related through the fluctuation dissipation relation to
appropriate static correlation functions.  A way to measure
such responses in the context of numerical simulations is 
the slave equation method which is used for Langevin type systems, 
and more precisely in the context of stochastic 
quantization \cite{pawu}. Via the static fluctuation
dissipation theorem the slave equation method can be used
to compute correlation functions and has  been successfully exploited 
in numerical simulations of quantum field theories and statistical 
spin systems \cite{phi4,spin}. 
The advantage of the slave
equation method is that it provides a way of measuring cumulant
correlators directly with properly estimated statistical errors. As
has been noted however \cite{phi4,spin}, there are circumstances in which
the slave equation method breaks down. These difficulties
are clearly illustrated by the simple model investigated in this
paper. Another advantage of the slave method is that the response can
be measured without imposing a small external field and thus 
unperturbed correlation functions can be measured simultaneously.  
Recently various numerical methods for measuring response functions in discrete
spin systems, without applying an external field,  have been 
proposed \cite{fred}.

Although not the subject of this paper, it is worth noting that for
systems in equilibrium more general fluctuation dissipation theorems exist
that relate dynamical response functions and dynamical correlation functions.
In  out of equilibrium systems that exhibit aging
various  types of  fluctuation dissipation theorems 
have been to shown to hold in mean field models and numerical 
and experimental evidence points to their validity in finite 
dimensional systems \cite{afdt}.  
Forms of fluctuation dissipation relations/theorems are also 
expected to hold in the steady state of certain non-equilibrium driven 
systems.

The model we investigate is the simplest possible, 
namely an over-damped particle 
moving in one dimension in a
background potential $\phi(x)$ subject to an external field $h$ and
thermal noise.  The Langevin equation for this system is
\begin{equation}
{\dot x}_t = -\phi'(x_t) + h + \eta_t~~,
\label{eqlang}
\end{equation}
where $\eta_t$ is zero mean Gaussian white noise with correlation
function
\begin{equation}
\langle \eta_t \eta_{t'}\rangle = 2T \delta(t-t')~~,
\end{equation}
with $T$ the temperature. At zero external field we define the
response function $\xi$ by
\begin{equation}
\xi_t = {\partial x_t \over \partial h}\vert_{h=0}
\end{equation}
Differentiating Eq.(\ref{eqlang}) with respect to $h$ and setting
$h=0$ we obtain the equation of motion of $\xi_t$ as
\begin{equation}
{\dot \xi}_t = -\phi''(x_t)\xi_t + 1~~ .
\label{eqslave}
\end{equation}
The variable $\xi_t$ is referred to as the slave and
Eq.(\ref{eqslave}) the slave equation corresponding to the process
$x_t$~. This terminology is obvious upon examining Eqs.(\ref{eqlang})
and (\ref{eqslave}) as we see that the process $\xi_t$ is driven by
the process $x_t$ but the evolution of $x_t$ is completely independent
of $\xi_t$~. Of course the analysis can be generalized to higher dimensions.
This is separately interesting and will be addressed in future work.

Processes whose  evolution is  similar to that given by  
Eq. (\ref{eqslave}) arise in a variety of physical contexts
such as the development of curvature in material line and surface elements
and magnetic fields transported by random flows 
\cite{pope,itd1,itd2,schekochihin}. Indeed the behavior exhibited by 
the probability density functions of the relevant slave variables is exactly 
as analyzed in our simple model. 
A related slave variable $\zeta_t$~,
\begin{equation}
\zeta_t={{\partial x_t}\over{\partial x_0}}~~,
\end{equation}
where $x_0$ is the initial value of $x_t$~, satisfies the slave equation
\begin{equation}
{\dot\zeta}_t= -\phi''(x_t)\zeta_t~~.
\label{eqslave2}
\end{equation}
Eq. (\ref{eqslave2}) is identical to Eq. (\ref{eqslave}) except for the
inhomogeneous term $+1$ on the right. Although the two equations are close
in form the presence of this inhomogeneous term radically affects the
statistics of $\xi_t$~. The behavior of the slave variable $\zeta_t$ is 
is a measure of the sensitivity of a system to its initial conditions
and thus related to Lyapounov exponents. It was recently shown how Lyapounov 
exponents could be analyzed  via supersymmetric quantum mechanics 
\cite{jorge}, our approach is not explicitly developed in terms of 
supersymmetry but it is clear that it could be rewritten in these terms.

In equilibrium in the presence of an  external field $h$~, the 
statistics of the process $x_t$ is described by the Gibbs-Boltzmann 
distribution
\begin{equation}
P_{GB}(x) = {1\over Z(h)}\exp\left(-\beta \phi(x)+ \beta hx \right)~~,
\label{eqgb}
\end{equation}
where
\begin{equation}
Z(h) = \int dx\ \exp(-\beta \phi(x) +\beta hx )
\end{equation}
is the canonical partition function for the process $x$~.  Clearly one
has, by linearity of an expectation of a probability distribution,
that
\begin{equation} 
\langle \xi\rangle_E = {\partial \over \partial h}\langle x\rangle_E~~,
\end{equation}
where the subscript $E$ denotes expectations taken in the equilibrium
state.  Using the form of the
Gibbs-Boltzmann distribution then yields the static
fluctuation-dissipation theorem
\begin{equation} 
\langle \xi\rangle_E = \beta\left( \langle x^2\rangle_E - \langle
x\rangle_E ^2 \right)\label{eqfdt}~~.
\end{equation}

In simulations, the mean of $\xi$ in equilibrium is easy to
determine. It therefore provides, up to a multiplicative factor, a
direct estimator for the variance of $x$~.  The numerical usefulness
of this result is two-fold. First a straight calculation of the
variance computed as in Eq.(\ref{eqfdt}) is not ideal from the 
the point of view of precision,  as it involves the subtraction of
two numbers, each potentially much larger than their
difference. Measuring $\xi$ avoids this difficulty. 
Second, from the variance of $\xi$ we have a proper
statistical estimate for the error in the susceptibility and the
variance of $x$~. The utility of the slave method hinges precisely on
the existence and size of the variance of $\xi$~, the slave variable.

Given the importance of $\xi$ as an estimator to the variance of $x$ 
by the fluctuation-dissipation relation and its direct physical
significance as a susceptibility, it is natural to investigate its
statistical properties and in particular its equilibrium distribution
function which we shall denote by $\rho(\xi)$~.  In this paper, we address 
this issue in detail.

The equilibrium probabilty density of $\xi$, $\rho(\xi)$~,  may be obtained
from the joint  probability density function (PDF) $P(x,\xi)$
via~,
\begin{equation}
\rho(\xi)=\int dx P(x,\xi)~~.
\end{equation}
The joint PDF satisfies \cite{risk}, in equilibrium
\begin{equation}
-H_{FP}P(x,\xi)+{\partial\over\partial\xi}\left(\left(\phi''(x)\xi-1\right)P(x,\xi)\right)
 =0~~,
\label{fpeq}
\end{equation}
where $H_{FP}$ is the forward Fokker Planck operator for the process
$x_t$ and is defined by
\begin{equation}
H_{FP} P = -T\pd2x P - \phi'(x) \pdx P - \phi''(x)P~~.
\end{equation}
The Gibbs-Boltzmann distribution for $x$, $P_{GB}(x)$~of Eq. (\ref{eqgb}), 
is recovered via
\begin{equation}
P_{GB}(x)=\int d\xi P(x,\xi)~~,
\end{equation}
and of course satisfies
\begin{equation}
-H_{FP}P_{GB}(x)=0~~.
\end{equation}

\section{Static fluctuation dissipation relation}

The fundamental nature of the static fluctuation dissipation
relation  between the  equilibrium susceptibility and the systems variance 
means it is illuminating to verify the relation 
Eq.(\ref{eqfdt}) from Eq.(\ref{fpeq} directly). We set
\begin{equation}
P(x,\xi)=P_{GB}(x)F(x,\xi)~~,
\end{equation}
and define $F_n(x)$~, where it exists, by the equation
\begin{equation}
F_n(x)=\int d\xi \xi^n F(x,\xi)~~.
\end{equation}
Clearly
\begin{equation}
F_0(x)=1~~,
\end{equation}
and
\begin{equation}
\langle\xi\rangle_E=\int dx\ P_{GB}(x)F_1(x)~~.
\end{equation}
It follows from Eq.(\ref{fpeq}) that
\begin{equation}
\left(T\pdx-\phi'(x)\right)\pdx
F(x,\xi)+\pdxi\left(\left(\phi''(x)\xi-1\right)F(x,\xi)\right)=0~~.
\end{equation}
If we multiply by $\xi$ and integrate over all $\xi$ we find
\begin{equation}
\left(T\pdx-\phi'(x)\right)\pdx F_1(x)-\phi''(x)F_1(x)+1=0~~.
\end{equation}
This can be rewritten in the form
\begin{equation}
\pdx\left(T\pdx F_1(x)-\phi'(x)F_1(x)+x\right)=0~~,
\end{equation}
which implies that
\begin{equation}
T\pdx F_1(x)-\phi'(x)F_1(x)+x-a=0~~,
\label{eqint}
\end{equation}
where $a$ is an integration constant. If we then multiply by $P_{GB}(x)$
and integrate over all $x$ we find
\begin{equation}
a=\langle x\rangle_E~~.
\end{equation}
We now multiply by Eq. (\ref{eqint}) by $(x-a)P_{GB}(x)$ and 
integrate over all $x$ to  obtain
\begin{equation}
-T\int dx\ P_{GB}(x)F_1(x)+\int dx\ P_{GB}(x)(x-a)^2=0~~,
\end{equation}
which is precisely the static fluctuation dissipation relation
Eq. (\ref{eqfdt}).
The derivation
therefore confirms the properties of the estimator $\xi_t$ as deduced
from the stochastic differential equation for $x_t$~.

\section{General Properties of the $\xi$-Probability Distribution}

It follows from Eq.(\ref{eqslave}) that for $\xi\simeq 0$~,
${\dot\xi_t}\simeq 1$~.  This implies that there is always a positive
flow of probability from negative to positive $\xi$~. In equilibrium
therefore the support for $P(x,\xi)$ lies in range $\xi>0$~. We shall
assume also, which can be justified subsequently, that $P(x,0)=0$~.

The Laplace transform of the joint probability function is
\begin{equation}
{\tilde P}(x,s)=\int_{0}^{\infty}d\xi e^{-s\xi}P(x,\xi)~~.
\end{equation}
It follows from our discussion that we expect ${\tilde
P}(x,s)\rightarrow 0$ faster than $s^{-1}$ as $s\rightarrow\infty$~,
and
\begin{equation}
\int_{0}^{\infty} d\xi e^{-s\xi}\pdxi P(x,\xi)=s{\tilde P}(x,s)~~.
\end{equation}

The large $\xi$ behavior of $P(x,\xi)$ is also of importance.  As
will become clear later this is strongly influenced by the behavior
of $\phi''(x)$~. Let there be an interval $U$ such that for $x\in U$~,
$\phi''(x) \simeq -g$ where $g>0$~. It follows from Eq.(\ref{eqslave}) while
$x$ remains in $U$ then $\xi$ will grow exponentially. If the time for
which $x$ remains in $U$ is $\tau$ then the excursion experienced by
$\xi$ will be roughly of the form
\begin{equation}
\xi=\xi_0e^{g\tau}~~,
\end{equation}
for some $\xi_0$~.  When $x$ leaves $U$~, $\xi$ will decay rapidly back
to small values.  Since the the process $x_t$ is essentially without
memory the distribution of $\tau$ will be exponential.
\begin{equation}
p(\tau)\simeq \mu\exp\left(-\mu\tau\right)~~.
\end{equation}
The large $\xi$ behavior of $\rho(\xi)$ is determined by these
excursions.  We have
\begin{equation}
\rho(\xi) \approx
{\mu\over{g\xi_0}}\left({\xi_0\over\xi}\right)^{1+\alpha^*}~~,
\label{eqpl}
\end{equation}
where $\alpha^*=\mu/g$~.  Hence we can expect a power law in the
distribution for large values of $\xi$ when there exists a region $U$
in which the potential $\phi(x)$ is concave.  Of course the argument
above does not allow an easy calculation or even estimation of the
exponent since the time spent in concave regions is affected by the
global structure of the potential.

\begin{figure*}
\includegraphics[width=.7\textwidth]{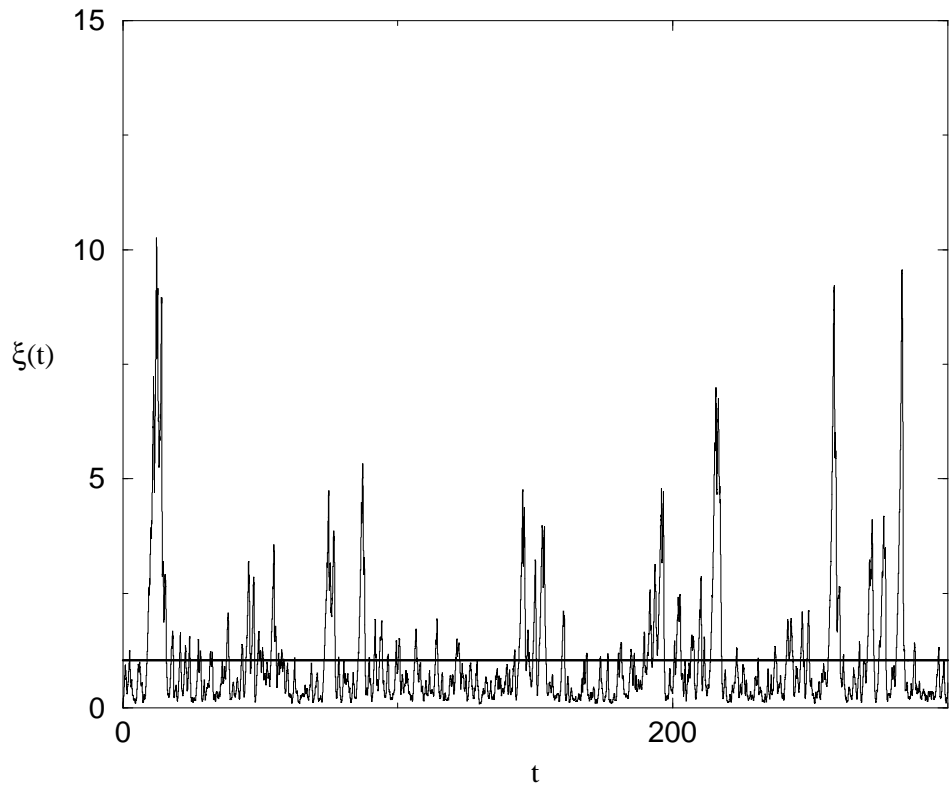}
\caption{Time series for slave $\xi_t$ with potential $\phi(x) =
(1-x^2)^2/4$ at $\beta = 1$~. Shown by the thick horizontal line is its
average value}
\label{figtsb1}
\end{figure*}

A potential which exhibits a region of concavity and for which
therefore we expect to find a power law behavior for $\rho(\xi)$~, is
$\phi(x) = (1-x^2)^2/4$~.  We have simulated the Langevin process
$x_t$ and its slave by integrating Eqs.(\ref{eqlang}) and
(\ref{eqslave}) using a second order stochastic Runga-Kutta method
\cite{rk}. Shown in Fig.  (\ref{figtsb1}) is the time series obtained
for $\xi$ at $\beta=1/T = 1$~. The mean value of the process as
predicted by the static fluctuation dissipation relation Eq.(\ref{eqfdt}) is
shown by the thick horizontal line.  A direct measurement of the time
series average confirms this result, as it should.

The most striking feature of the time series however is that $\xi_t$
spends most of the time below the average value with
intermittent spikes rising to large values. These spikes are the means
by which the time series fills the power law tail in the distribution
for $\rho(\xi)$ as indicated by the intuitive argument explained
above. The implication of this result for simulations is that it is
essential to include the intermittent upward excursions if correct
estimates are to be obtained for the susceptibility.  If therefore one
were to measure $\langle \xi\rangle_E$ by taking a time series average,
then in order to obtain a satisfactory estimate one must ensure that
the time interval of the measurement is of a length sufficient to
sample the rare but large excursions that represent the power law tail
of $\rho(\xi)$~.  The typical value of $\xi$ that one measures during
a simulation is well below the mean value. Thus measuring over too
short a time scale, or for other reasons omitting the large excursions
will lead to one to an under estimate of $\langle\xi\rangle_E$~.

Another approach to the fluctuation dissipation theorem would be to
be to compute numerically or measure physically the expectation value 
$\xi_a$ of the slave variable $\xi$ and the variance of the the original 
variable $x$ and use these results to provide an estimate $T_{eff}$
of the temperature of the system.
\begin{equation}
T_{eff} = {\langle x^2 \rangle_E - \langle x\rangle_E^2 \over \xi_a}~~.
\end{equation}
If for the reasons discussed above we under estimate $\xi_a$
then our estimate for the temperature will be too high,
\begin{equation}
T_{eff}\ge T~~.
\end{equation}
It is interesting to note that this result with $T_{eff}/T \ge 1$
is also seen in aging systems where the dynamical fluctuation
dissipation theorem is violated because the system is not in
equilibrium \cite{eftemp}.  However here the apparent violation is due to
an error of measurement in an equilibrium state and hence of 
somewhat different origin.

The value of the exponent $\alpha^*$ appearing in  Eq.  (\ref{eqpl})
may be obtained from the numerical simulation by fitting a straight line 
to the large $\xi$ region of the log-log plot of the numerically generated
histogram of $\xi$~.
We have carried out this procedure for a range of values of the
temperature $T$ and the results are shown in
table (\ref{t1})~. Most importantly we see that 
the exponent $\alpha^*$ depends continuously on temperature.
In addition, two features of the behavior of $\alpha^*$  
in these numerical results stand out. First,
as $T\to \infty $ then it appears $\alpha^* \to
\infty$~ (as $\alpha^*$ increases an accurate fit of the power law
tail's exponent becomes difficult due to the lack of statistical weight
in the tail). Second, as $T \to 0 $  the
numerics is consistent with $\alpha^* \to 1$~. This latter result is 
particularly significant. At
low temperature the PDF for $x$ takes the form
\begin{equation}
P_{GB}(x)={1\over 2}(\delta(x-1)+\delta(x+1))~~.
\end{equation}
This implies that $\langle x\rangle_E=0$ and $\langle
x^2\rangle_E=1$~. It follows from the static fluctuation dissipation relation 
that
\begin{equation}
\langle\xi\rangle_E\simeq {1\over T}~~,~~~~~\mbox{as}~~T\rightarrow
0~~.
\end{equation}
For large $\xi$ we have
\begin{equation}
\rho(\xi)\simeq {C\over\xi^{1+\alpha^*}}~~,\label{defc}
\end{equation}
for some positive $C$~. It follows that
\begin{equation}
\langle\xi\rangle_E=\int^\infty d\xi\xi\rho(\xi) \simeq\int^\infty
      d\xi\xi{C\over\xi^{1+\alpha*}}\simeq{C\over(\alpha^*-1)}~~.
\end{equation} 
Comparing this result with the static fluctuation dissipation relation we see
that at low temperature we should expect $\alpha^*\to
1$~ and more specifically 
$C/(\alpha^*-1) \approx 1/T$~. 
This is consistent with the numerical results shown in table (\ref{t1})~.
 
For a given value of $\alpha^*$ the existence of a power law tail for
$\rho(\xi)$ means that moments $\langle\xi^n\rangle_E$ diverge for
$n>\alpha^*$~. At low $T$, where $\alpha^*<2$, therefore even the
variance of the estimator for the susceptibility has become
divergent. At this point it has ceased to be a useful estimator and
provides no reasonable estimate for a statistical error on the
susceptibility.  The slave equation method therefore becomes
ineffective under these circumstances.  These issues in relation to
simulations of quantum field theory and spin models in statistical
mechanics have been noted before \cite{phi4,spin}.

\btab
\btabu{|c|c|}\hline
$T$ & $ \alpha^*$(numerics)  \\\hline
0.66 & 2.46(3)\\\hline
0.50 & 1.86(2) \\\hline
0.40 & 1.67(1) \\\hline
0.33 & 1.54(1)  \\\hline
\etabu
\caption{\label{t1}  The exponent $\alpha^*$~, estimated by straight
line fit to the tail of the log-log plot of the numerically generated 
histogram, as a function of $\beta$ for the potential $\phi(x) =
(1-x^2)^2/4$~. }
\etab

\section{General Theory}

The above observations on the nature of the PDF of $\xi$
are much clarified and rendered more robust by an understanding of
the general theory of the joint PDF. In order to pursue the analysis it
is convenient to make a standard transformation that renders $H_{FP}$
into self adjoint form.  We define $Q(x,\xi)$ so that
\begin{equation}
P(x,\xi) = Q(x,\xi) {\exp(-\beta \phi(x)/2)\over Z^{1\over 2}}~~.
\end{equation}
It follows that $Q$ obeys
\begin{equation}
-H_0 Q(x,\xi) + \pdxi\left( (\xi \phi''(x)
-1) Q(x,\xi)\right) = 0~~,
\end{equation}
where the manifestly self adjoint operator $H_0$ is given by
\begin{equation}
H_0 = -T \pd2x + V(x)
\end{equation}
with
\begin{equation}
V(x)=\left( {1\over 4 T} \left(\phi'(x)\right)^2 - {1\over 2}
\phi''(x)\right)~~.
\end{equation}
We introduce the Laplace transform of $Q$ with respect to the variable
$\xi$
\begin{equation}
{\tilde Q}(x,s) = \int_0^\infty d\xi\ \exp(-s\xi) Q(x,\xi)~~,
\end{equation}
and thus corresponding Laplace transform of $P$ is given by
\begin{equation}
{\tilde P}(x,s) = {\exp\left(-\beta \phi(x)/2\right) {\tilde Q}(x,s)
\over Z^{1\over2}}~~.
\end{equation}
We note that
\begin{equation}
{\tilde P}(x,0) = {\exp\left(-\beta \phi(x)\right)\over Z}~~,
\end{equation}
which thus gives us the initial conditions for ${\tilde Q}$ at $s=0$
to be
\begin{equation}
{\tilde Q}(x,0) = {\exp\left(-\beta \phi(x)/2\right)\over Z^{1\over
2}}~~.
\label{eqqic}
\end{equation}
The evolution equation for ${\tilde Q}$ is
\begin{equation}
H_0 {\tilde Q} +s{\tilde Q} + s\pds {\tilde Q} = 0 \label{eqqt}~~.
\end{equation}
We notice that a particular solution of Eq. (\ref{eqqt}) can  be written
as a power-series expansion in $s$
\begin{equation}
{\tilde Q}^{(\alpha)}(x,s) = \sum_{n=0}^\infty s^{\alpha+n}\
f_n^{(\alpha)}(x)~~,
\end{equation}
where the indicial equation determining the allowed values of $\alpha$
is
\begin{equation}
H_0 f_0^{(\alpha)} + \alpha \phi''(x) f_0^{(\alpha)} = 0~~.
\label{eqie}
\end{equation}
The general solution may then be written as a linear superposition of
these particular solutions:
\begin{equation}
{\tilde Q}(x,s) = \sum_{\alpha} w_\alpha {\tilde Q}^{(\alpha)}(x,s)~~.
\label{eqqd}
\end{equation}
We notice that $\alpha = 0$ is always a solution of Eq. (\ref{eqie})
with a corresponding $f_0^{(0)}$ given by
\begin{equation}
f_0^{(0)}(x) ={\exp\left(-\beta \phi(x)/2\right)\over Z^{1\over 2}}~~.
\end{equation}
The initial condition Eq. ({\ref{eqqic}) means that $w_\alpha = 0$ for
$\alpha < 0$~, otherwise ${\tilde Q}$ would diverge at $s=0$~.  The
solution must therefore be of the form
\begin{equation}
{\tilde Q}(x,s) = \sum_{n= 0}^\infty s^n\ f_n^{(0)}(x)+ \sum_{\alpha >
0} w_\alpha {\tilde Q}^{(\alpha)}(x,s) \label{eqqf}~~.
\end{equation}
It is easy to see, on reconstructing $\rho(\xi)$ from the above
Laplace transform, that it must have the form
\begin{equation}
\rho(\xi)\propto {1\over \xi^{1+\alpha^*}} \label{eqpl2}
\end{equation}
for large $\xi$ where $\alpha^*$ is the smallest strictly positive
solution to the indicial equation with $w_{\alpha^*} \neq 0$~. If there
are no solutions to the indicial equation with $\alpha > 0$ then
${\tilde Q}(x,s)$ must be given by
\begin{equation}
{\tilde Q}(x,s) = \sum_{n= 0}^\infty s^n\ f_n^{(0)}(x)
\end{equation}
and is analytic in $s$ and all of the moments of $\xi$ will presumably
exist; thus $\rho(\xi)$ will not have a power law tail at large
$\xi$~.  The physical arguments leading to Eq. (\ref{eqpl}) suggest
that Eq. (\ref{eqpl2}) should hold when there is a region where
$\phi''(x) < 0$~. In addition the same physical argument and the static
fluctuation dissipation relation also show that we should have
$\alpha^* > 1$ for finite $\beta$~, otherwise $\xi$ will diverge which means
that the variance of $x$ on the right hand side of Eq. (\ref{eqfdt})
diverges, which is clearly not possible for a sufficiently
confining potential. We shall now confirm this physical picture
mathematically.  If we define the operator
\begin{equation}
A_0 = \sqrt{T} \pdx + {1\over 2\sqrt{T}} \phi'(x)~~,
\end{equation}
then clearly we may write $H_0 =A_0^\dagger A_0$ which shows that
$H_0$ is positive semi-definite. If $f$ is a solution to the indicial
equation Eq. (\ref{eqie}), multiplying by $f$ and integrating over all
$x$ gives
\begin{equation}
\int dx\ fH_0 f + \alpha \int dx\ f^2 \phi''(x) = 0~~.
\end{equation}
When $\phi''(x) \ge 0$ for all $x$~, it follows that $\alpha \leq
0$~. As expected, no power law behavior is possible in this case.

The indicial equation Eq. (\ref{eqie}) may also be written as
\begin{equation}
-T \pd2x f + \left( {(1-2\alpha)^2\over 4 T} \left(\phi'(x)\right)^2 -
{1-2\alpha \over 2} \phi''(x)\right) f + {\alpha (1-\alpha)\over T}
\left(\phi'(x)\right)^2 f = 0~~,
\end{equation}
which is equivalent to
\begin{equation}
H_\alpha f + {\alpha (1-\alpha)\over T}\left(\phi'(x)\right)^2 f = 0~~,
\end{equation}
where the positive semi-definite $H_\alpha$ is given by
\begin{equation}
H_\alpha = A_\alpha^\dagger A_\alpha~~,
\end{equation}
with
\begin{equation}
A_\alpha = \sqrt{T} \pdx + {1-2\alpha\over 2\sqrt{T}} \phi'(x)~~.
\end{equation}
This thus yields
\begin{equation}
\int dx\ fH_\alpha f + {\alpha(1-\alpha)\over T} \int dx\ f^2
\left(\phi'(x)\right)^2 = 0~~,
\end{equation}
which implies that $\alpha(1-\alpha) < 0$~, thus if $\alpha$ is
positive then we must have $\alpha >1$~, again confirming the physical
reasoning of the Introduction.

We shall now show that if there exists a region where $ \phi''(x) <
0$~, then there is a solution to the indicial equation with $\alpha
>0$, and from the preceding argument, if such an $\alpha$ exits then
$\alpha >1$~.

We consider the following eigenvalue equation, 
\begin{equation}
H_0 \psi_0 + \alpha \phi''(x) \psi_0 =
E_0(\alpha)\psi_0~~,
\end{equation}
where $E_0$ denotes the ground
state energy and $\psi$ the corresponding ground state wave function.
Consider this eigenvalue problem at $\alpha =0$~, here we have
\begin{eqnarray}
\psi_0\vert_{\alpha=0} &=& {1\over Z^{1\over 2}}\exp\left(-\beta
\phi(x)/2 \right) \\ E_0(0) &=& 0~~.
\end{eqnarray}
First order perturbation theory shows us that
\begin{eqnarray}
{\partial \over \partial \alpha}E_0(\alpha) &=& \int dx\ \psi_0^2
\phi''(x) \nonumber \\ &=& {1\over ZT} \int dx \left(\phi'(x)\right)^2
\exp\left(-\beta \phi(x) \right)~~.
\end{eqnarray}
Thus at $\alpha = 0$ we have ${\partial \over \partial
\alpha}E_0(\alpha) > 0$~, thus there is a region of $\alpha >0$ where
$E_0(\alpha) > 0$ and there can be no acceptable solution $\alpha$ to
the indicial equation in that region.

We shall now use the well known variational formula
\begin{equation}
E_0(\alpha) = {\rm min}_{\psi}\ \int dx \left[ T \left({\partial\psi
\over \partial x}\right)^2 + \left( {1\over 4T} \left(\phi'(x)\right)^2 + 
{1\over 2}(1-2\alpha)\phi''(x)\right) \psi^2 \right] \label{eqvar}~~,
\end{equation}
where the minimum is taken over all functions such that $\int dx\ 
\psi(x)^2 = 1$~.  Define $g = - {\rm min}_x\{\phi''(x)\}$ and consider
the case where $g>0$ and let $x_0$ be a point where this minimum is
obtained. Without loss of generality we take $x_0=0$~. Now consider the
trial wave function
\begin{equation}
\psi^*(x) = \left(c\over 2 \pi\right)^{1\over 4} \exp(-cx^2/4)
\end{equation}
for $c$ large and positive. Using Eq. (\ref{eqvar}) we obtain
\begin{equation}
E_0(\alpha) \leq {Tc\over 4} + \left({1\over
4T}\left(\phi'(0)\right)^2  - {g\over 2}(2\alpha -1)\right)
+ O\left({1\over \sqrt{c}}\right)~~.
\end{equation}
Thus if $c\gg 1$ and $\alpha \gg c$ then we
have $E_0(\alpha) < 0$~.  Assuming the continuity of $E_0(\alpha)$~,
along with the fact that $E(\alpha)$ is positive in a region $(0, l)$
for some $l>0$~, we have shown the existence of $\alpha^* >0$ such that
$E(\alpha^*) = 0$~.

Now it remains to be shown that in the case where $\phi(x)$ is concave
for some range of $x$~, that the coefficient $w_{\alpha^*}$ in
Eq. (\ref{eqqd}) is non-zero. We recall the boundary condition
$P(x,0)=0$ which means that for large $s$ that ${\tilde Q}(x,s)$ must
decay more quickly than $1/s$ at large $s$~.

The Eq. (\ref{eqqt}) may be written as
\begin{equation}
{\tilde Q}(x,s) = -\int dx'\ G(x,x';s) s\pds {\tilde Q}(x',s)
\phi''(x')~~,
\end{equation}
where $G$ is the Green's function obeying
\begin{equation}
(H_0 + s) G(x,x';s) = \delta(x-x') \label{eqgf}~~.
\end{equation}
For fixed $x$ and $x'$ and $s \gg1$ we have from Eq. (\ref{eqgf})
\begin{equation}
G(x,x';s) \approx {\delta(x-x')\over s}~~,
\end{equation}
which means that for large $s$
\begin{equation}
{\tilde Q}(x,s) \approx -\phi''(x) \pds {\tilde Q}(x,s)
\end{equation}
and hence
\begin{equation}
{\tilde Q}(x,s) \approx H(x) \exp\left(-s/
\phi''(x)\right)~~. \label{eqexp}
\end{equation} 
This means that in particular if there is a point $x$ where $\phi$ is
concave then ${\tilde Q}^{(0)}(x,s)$ diverges there and thus the full
solution needs to have at least one $w_\alpha \neq 0$ as ${\tilde
Q}^{(\alpha)}(x,s)$ has the same divergent behavior there as $s \to
\infty$~, the coefficients must then be chosen to cancel the
divergence. A similar mechanism was identified in simplified discrete 
versions of the the slave equation of the model discussed in this paper \cite{ITD3}.

\section{Specific Examples}

\subsection{The Simple Harmonic Oscillator}
The simplest example one can consider is the simple harmonic
oscillator with
\begin{equation}
\phi(x) = {\lambda x^2\over 2}~~.
\end{equation}
From the theory of the precedent section we know that there should be
no power law behavior in $\rho(\xi)$~, and although the problem can be
explicitly solved it is instructive to work through the
mathematics as formulated in Section III.

The indicial equation in this case is
\begin{equation}
-T \pd2x f + {\lambda^2 x^2\over 4T}f = {\lambda\over 2}(1-2\alpha) f~~.
 \label{sho}
\end{equation}
The left hand side of Eq. (\ref{sho}) is the Hamiltonian for a quantum
simple harmonic oscillator of mass $m=1/2T$ and frequency $\omega =
\lambda$~. The energy levels are thus $E_n = (n+{1\over 2})\lambda$ and
comparing with the right hand side of Eq. (\ref{sho}) immediately
yields that the solutions for $\alpha$ are $\alpha = -n$ and hence, as
predicted,  are all negative. Clearly the slave equation reads
\begin{equation}
{\dot \xi}_t = -\lambda \xi + 1~~,
\end{equation}
and so the equilibrium distribution of $\xi$ is a delta function at the fixed
point $\xi = 1/\lambda$~.
\begin{equation}
\rho(\xi) = \delta(\xi -{1\over \lambda})~~.
\end{equation}

\subsection{The Potential $\phi = |x|$} 
We now consider the potential
\begin{equation}
\phi(x) = |x|~~.
\end{equation}
Again there can be no power law behavior. The interesting point about
this potential is that the full distribution of $\xi$ can be
computed. One can easily solve for ${\tilde P}(x,s)$ to obtain
\begin{equation}
{\tilde P}(x,s) = {\lambda(s)\over 2} \exp\left(-\lambda(s) |x|\right)
\exp\left(-2(\beta s + {1\over 4})^{1\over 2}\right)~~,
\end{equation} 
where
\begin{equation}
\lambda(s) = {\beta + (\beta^2 + 4\beta s)^{1\over 2}\over 2}~~.
\end{equation}
Then after integrating over $x$~, the Laplace transform can be
inverted to yield
\begin{equation}
\rho(\xi) = {\beta e\over \pi^{1\over 2} (\beta \xi)^{3\over 2}}
\exp\left( -{\beta \xi\over 4} -{1\over \beta
\xi}\right)~~. \label{pdxiv}
\end{equation}
Notice that the large $s$ behavior is not that predicted by
Eq. (\ref{eqexp}) due to the delta function singularity in $\phi''(x)$
at $x=0$~, where this occurs it is easy to see that one has a behavior
of the form $\exp(-A \sqrt{s})$ but the conclusions stay the
same. Indeed the simplest cases exhibiting a power law distribution in
$\xi$ are those where $\phi''(x)$ is composed of delta functions, 
as is the case for continuous piecewise linear or quadratic
potentials $\phi$~. 

\subsection{The piece-wise continuous quadratic potential}
Here we consider the potential
\begin{equation}
\phi(x) = {1\over 2}(|x|-1)^2 - h x \label{phic}~~.
\end{equation}
In this case we again expect a power law tail in the distribution of
$\xi$~.  The indicial equation in this case is
\begin{eqnarray}
-T \pd2x f + \left( {1\over 4T}(x-h-1)^2 -{1\over 2}(1-2\alpha)\right)f &=&
0 \ ;x>0 \nonumber \\ -T \pd2x f + \left( {1\over 4T}(x-h +1)^2 -{1\over
  2}(1-2\alpha)\right)f &=& 0 ;x<0
\end{eqnarray}
along with the continuity of $f$ at $x=0$ and the jump condition.
\begin{equation}
-T\left( {\partial f\over \partial x}|_{0^+} -
{\partial f\over \partial x}|_{0^-}
\right) + (1-2\alpha) f(0) = 0~~.
\end{equation}
The solution which decays as $|x| \to \infty$ is
\begin{eqnarray}
f(x)&=& A_+ D_{-\alpha}\left(\sqrt{\beta}(x-h-1)\right) \ ;\ x\geq0
\nonumber \\ f(x) &=& A_- D_{-\alpha}\left(\sqrt{\beta}(h-1-x)\right)
\ ;\ x\leq 0~~,
\end{eqnarray}
where $D_p$ denotes a parabolic cylinder function of index $p$
\cite{grad}.  Using the jump condition and continuity at $x=0$~, along
with the identity \cite{grad}
\begin{equation}
D'_p(z) -{z\over 2}D_p(z) +D_{p+1}(z) = 0~~,
\end{equation}
we find that $\alpha$ obeys the equation
\begin{equation}
\Gamma(\alpha,\beta,h)= 0 \label{eqpb}
\end{equation}
where
\begin{eqnarray}
\Gamma(\alpha ,\beta ,h)&=& 2 \sqrt{\beta}(\alpha-1)
D_{-\alpha}(-\sqrt{\beta}(1+h)) D_{-\alpha}(-\sqrt{\beta}(1-h))
\nonumber \\ &-& D_{1-\alpha}(-\sqrt{\beta}(1-h))
D_{-\alpha}(-\sqrt{\beta}(1+h)) - D_{1-\alpha}(-\sqrt{\beta}(1+h))
D_{-\alpha}(-\sqrt{\beta}(1-h))~~.
\end{eqnarray}

Numerically solving Eq. (\ref{eqpb}) shows that we have one positive
root $\alpha^*$~, the root $\alpha = 0$ while all the others are
negative.  An example is shown in Fig (\ref{figpara})

\begin{figure*}
\includegraphics[width=.7\textwidth]{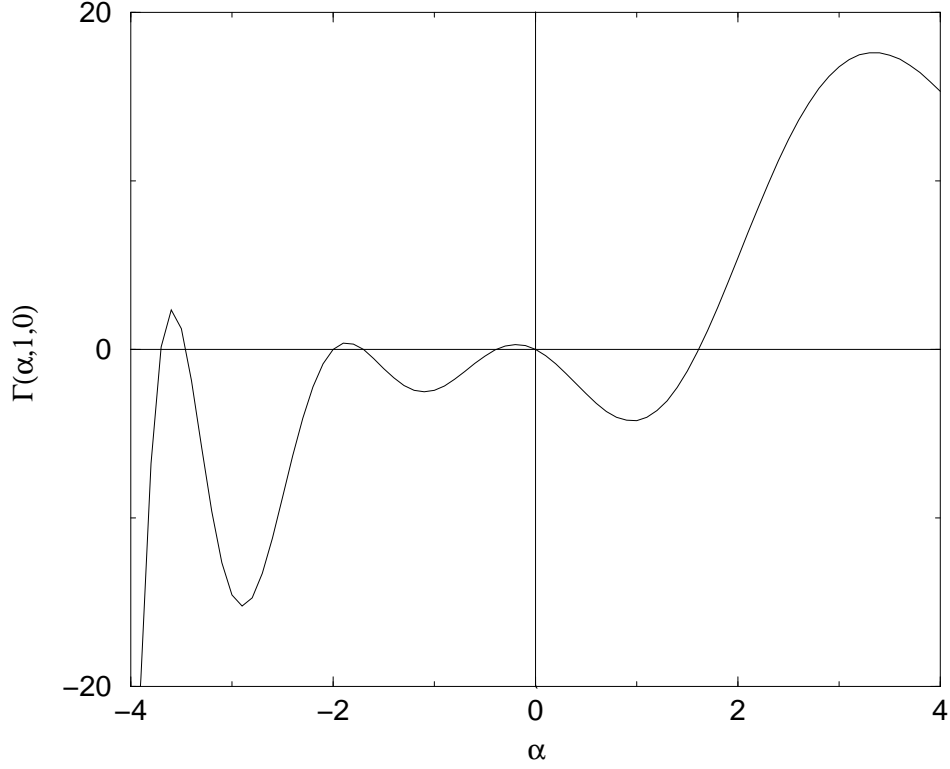}
\caption{Graph of $\Gamma(\alpha,\beta,h)$ against $\alpha$ at
$\beta=1$ and $h=0$~. Note there is only one strictly positive solution
to $\Gamma(\alpha,\beta,h)=0$~. }
\label{figpara}
\end{figure*}
At small $\beta$ we can show that
\begin{equation}
\alpha^*\approx 1/\beta~~,
\end{equation}
while the large $\beta$ behavior depends on the value of $h$~. In the 
case $|h|<1$ the system has two local minima and we find that
as $T\to 0$ then $\alpha^*$ is  given by
\begin{equation}
\alpha^* \approx 1 + {1\over \sqrt{8\pi \beta}}[\exp(-\beta(1+h)^2/2)+
\exp(-\beta(1-h)^2/2)] \label{a1}~~.
\end{equation}
Here we see that in the zero temperature limit $\alpha^* \to 1$~. 
In addition the asymptotic form for $\alpha^*$ Eq. (\ref{a1}) 
tells us that the coefficient $C$ of the power law tail must behave as 
\begin{equation}
C\sim \sqrt{{\beta \over 2\pi}}\exp(-\beta/2)
\end{equation}
in the case when $h=0$~. This follows from the static fluctuation 
dissipation theorem and the fact that the variance of $x$ is 
non-zero as $T\to 0$ when $h=0$~. 
Thus although the power law exponent $\alpha^*$ decreases, the coefficient
of the power law component of the PDF is, in this case, tending to zero
exponentially quickly. This can be understood physically as the excursions
into the region where $\phi$ is concave are into regions of high energy,
whose Boltzmann weight is exponentially suppressed.

In the case $|h| >1$ there is only one local minimum and we find 
as $T\to 0$ that 
\begin{equation}
\alpha^* \approx 1 + {|h|-1\over 2}~~.
\end{equation}

The predictions obtained by our method may be confirmed by numerical 
simulations. In the numerical simulations one needs $\phi'$ to be continuous 
and so the cusp in the potential at $x=0$ needs to be regularized. We take a
small interval $[\epsilon,\epsilon]$ where we set $\phi'(x) = (\epsilon -1)x 
/\epsilon + h$~, this choice ensures  $\phi'(x)$ is continuous and in the limit
$\epsilon \to 0$ we recover the  potential of Eq. (\ref{phic}).
Simulation results are performed with $\epsilon = 0.05$ and 
compared with the analytical result for the potential $\phi$ of 
Eq. (\ref{phic}). A table comparing the exponents obtained from the numerical 
simulation and from solving Eq. (\ref{eqpb}) for  the case $h=0$ 
is shown in table (\ref{t2}), the  agreement is excellent and the deviation 
is compatible with it being of order $\epsilon = 0.05$~. Similar agreement is
found when $h$ is non-zero.

\btab
\btabu{|c|c|c|}\hline
$T$ & $ \alpha^*$(numerics) &~~$\alpha^*$ (predicted) \\\hline
2.0 & 2.54(1) & 2.560 \\\hline
1.0 & 1.61(4) & 1.612 \\\hline
0.5 & 1.22(1)&  1.191 \\\hline
0.4 & 1.12(1)&  1.121 \\\hline
\etabu
\caption{\label{t2} Exponent $\alpha^*$ evaluated from histogram
of $\xi$ numerically generated for the potential $\phi$ of Eq. (\ref{phic})
regulated at the origin with $\epsilon = 0.05$ compared with analytical
prediction of Eq. (\ref{eqpb}) for various values of $T$ and at $h=0$~.}
\etab

\subsection{The `W' Potential}
Here we consider the `W' shaped potential given by
\begin{eqnarray}
\phi(x) &=& a |x-1|~~~~~~~~|x| \leq 1 \nonumber \\ &=& |x| -1~~~~~~~~~~|x| \geq
1~~.
\end{eqnarray}
The second derivative of the potential is thus given by
\begin{equation}
\phi''(x) = -2a \delta(x) +
(1+a)\delta(x-1) + (1+a)\delta(x+1)~~,
\end{equation}
and so from the general theory of section III we expect a power law
behavior where $a> 0$ and $a < -1$~. The indicial equation is given by
\begin{eqnarray}
-T \pd2x f + {a^2\over 4T}f &=& 0 \ \ |x| \leq 1 \nonumber \\ - T \pd2x
f + {1\over 4T}f &=& 0 \ \ |x| \geq 1~~,
\end{eqnarray}
along with the jump conditions
\begin{eqnarray}
-2T \pdx f|_{0} + a(1-2\alpha)f(0) &=& 0 \\ T \left[ \pdx f|_{1^+} 
-\pdx f|_{1^-}\right] - ({1\over2}-\alpha)(1+a)f(1) &=&0~~.
\end{eqnarray}
The solution of the indicial equation, which decays as $|x| \to
\infty$ is
\begin{eqnarray}
f(x) &=& A_+\exp(\beta a |x|/2) + A_-\exp(-\beta a |x|/2)~~~~~~ |x| \leq
1 \nonumber \\ &=& B \exp(-\beta |x|/2)~~~~~~~~~~~~~~~~~~~~~~~~~~~~~~~~~~|x| \geq 1~~.
\end{eqnarray}
Continuity at $x=1$ then gives
\begin{equation}
B \exp(-\beta /2) = A_+\exp(\beta a/2) + A_-\exp(-\beta a /2)~~,
\end{equation}
allowing for the elimination of the variable $B$~. The vector
\begin{equation}
{\bf u} = \pmatrix{A_+ &\cr A_-&\cr}
\end{equation}
is then determined by $M{\bf u} = 0$~, where
\begin{equation}
M = \pmatrix{-\alpha &1-\alpha \cr \exp(\beta a/2)\alpha (1+a)\ &
\exp(-\beta a/2)(\alpha (1+a) -a)\cr}~~.
\end{equation}
The possible values of the exponent $\alpha$ are then determined 
by the existence of a solution such that
${\bf u}\neq 0$~, that is to say ${\rm det} M =0$~, which yields the
solution $\alpha = 0$ or
\begin{equation}
\alpha =\alpha^* = 1 + {1\over (1+a)\left(\exp(\beta a)-1\right)}~~.
\label{eqaa}
\end{equation}
We see from Eq. (\ref{eqaa}) that $\alpha^*$ is positive and greater
than one in the region where $\phi$ has a concave component as
predicted by the general theory. In the region $a>0$ we find that
\begin{eqnarray}
\alpha^* \approx 1 + {1\over a(1+a)\beta} \ \ {\rm as } \ \beta \to
0 \nonumber \\ \approx 1 + {\exp(-\beta a)\over (1+a)} \ \ {\rm as } \
\beta \to \infty~~.
\end{eqnarray} 
In the region $a<-1$ we find
\begin{eqnarray}
\alpha^* \approx 1 + {1\over |a| |1+a|\beta} \ \ {\rm as } \ \beta
\to 0 \nonumber \\ \approx 1 + {1\over |1+a|} \ \ {\rm as } \ \beta
\to \infty~~.
\end{eqnarray}

We see in the case $a<-1$ in the limit $T\to 0$ that $\alpha^* >1$~, this
must be the case from the static fluctuation dissipation relation: 
here  there is  only one minimum and we therefore have
\begin{equation}
\beta \langle x^2 \rangle_E \approx {\beta \int dx \ x^2 \exp(-\beta|a| |x|)
\over   \int dx \exp(-\beta |a| |x|)} \approx {2T\over a^2}~~, 
\end{equation}
and hence $\langle \xi\rangle_E$ has no divergence as $T\to 0$~.

An interesting point emerges here. If the variance of $x$ is non-zero
in the limit $T\to 0$  the static fluctuation dissipation theorem
tells us that the coefficient $\alpha^* \to 1$ in this limit. However,
even if the variance of $x$ tends  to zero then 
$\alpha^*$ may still tend to one
in the zero temperature limit.  This is seen in the case of the potential 
$\phi(x) = (|x|-1)^2/2 - hx$ where the minima are only degenerate at 
$h=0$~. The average value of $\xi$ stays finite because the prefactor, 
denoted in this paper by $C$~, of the  power law component of the 
PDF of $\xi$ is tending to zero sufficiently rapidly. In the cases
we have examined here it seems that $\alpha^*\to 1$ in the zero temperature
limit when there are at least two  local minima of the potential $\phi$~.
It seems possible therefore that the zero temperature behavior of the 
exponent $\alpha^*$ encodes geometrical or topological properties of the
potential $\phi$~. 

\section{Conclusions}
We have analyzed the equilibrium distribution of a slave variable 
$\xi_t$ which is the estimator of the susceptibility of a 
one dimensional Langevin process $x_t$ . Even though the equilibrium statistics
of $x_t$ are such that  all moments are finite (for a sufficiently confining
potential), the probability density function of the slave $\xi_t$ can have 
power law tails characterized by a temperature dependent exponent. This
power law behavior is present when the potential $\phi$ has a concave 
component. The origin of this power law can be understood from simple
qualitative arguments. The behavior of the power-law exponent can be analyzed
in the steady state using the Fokker-Plank equation for the 
equilibrium joint probability density function $P(x,\xi)$ for $(x_t,\xi_t)$~.
The exponent is large at high temperatures and decreases on decreasing the 
temperature. As the temperature is reduced its higher order moments 
of the slave $\xi_t$ diverge and ultimately the variance of the 
slave may diverge thus rendering it a poor
estimator for the  susceptibility. This pathology in the slave statistics 
had been observed in Langevin simulations of spin and quantum systems 
\cite{phi4,spin}. A number of exactly soluble cases were analyzed and the
results confirmed by numerical simulation.

In future work it would be interesting to generalize our results to higher
dimensional systems, notably interacting systems where phase transitions
may occur. The temporal evolution  of the PDF of
the slave is also worthy of future study. 
It would be interesting to know how quickly the 
tails of the slave's PDF fill out and after what time it becomes equilibrated.
One would also like to understand over which timescale temporal averages need
to be carried out in order to numerically verify the static fluctuation
dissipation relation.  Finally the analysis developed here could prove useful 
in the analysis of similar slave variables occurring in Langevin systems.

\pagestyle{plain}

\end{document}